\documentclass[sigconf]{acmart}

\AtBeginDocument{%
  \providecommand\BibTeX{{%
    \normalfont B\kern-0.5em{\scshape i\kern-0.25em b}\kern-0.8em\TeX}}}

\setcopyright{acmcopyright}
\copyrightyear{2018}
\acmYear{2018}
\acmDOI{XXXXXXX.XXXXXXX}

\acmConference[Conference acronym 'XX]{Make sure to enter the correct
  conference title from your rights confirmation emai}{June 03--05,
  2018}{Woodstock, NY}
\acmPrice{15.00}
\acmISBN{978-1-4503-XXXX-X/18/06}

\usepackage{caption}
\usepackage{graphicx}
\usepackage{subcaption}
\usepackage{makecell}
\usepackage[all]{nowidow}
\usepackage[ruled,linesnumbered]{algorithm2e}

\begin{document}
\title{[Experiments \& Analysis] \\ Deep Clustering for Data Cleaning and Integration}

\author{Hafiz Tayyab Rauf} 
\orcid{0000-0002-1515-3187}
\affiliation{%
  \institution{Department of Computer Science, University of Manchester}
  \city{Manchester}
  \country{UK}
  \postcode{M13 9PL}
}
\email{hafiztayyab.rauf@manchester.ac.uk}

\author{Andre Freitas}
\orcid{0000-0002-4430-4837}
\affiliation{%
  \institution{Department of Computer Science, University of Manchester}
  \city{Manchester}
  \country{UK}
  \postcode{M13 9PL}
}
\email{andre.freitas@manchester.ac.uk}

\author{Norman W. Paton}
\orcid{0000-0003-2008-6617}
\affiliation{%
  \institution{Department of Computer Science, University of Manchester}
  \city{Manchester}
  \country{UK}
  \postcode{M13 9PL}
}
\email{norman.paton@manchester.ac.uk}

\begin{abstract}
Deep Learning (DL) techniques now constitute the state-of-the-art for important problems in areas such as text and image processing, and there have been impactful results that deploy DL in several data management tasks. Deep Clustering (DC) has recently emerged as a sub-discipline of DL, in which data representations are learned in tandem with clustering, with a view to automatically identifying the features of the data that lead to improved clustering results. While DC has been used to good effect in several domains, particularly in image processing, the impact of DC on mainstream data management tasks remains unexplored. In this paper, we address this gap by investigating the impact of DC in data cleaning and integration tasks, specifically {\it schema inference}, {\it entity resolution}, and {\it domain discovery}, tasks that represent clustering from the perspective of tables, rows, and columns, respectively. In this setting, we compare and contrast several DC and non-DC clustering algorithms using standard benchmarks. The results show, among other things, that the most effective DC algorithms consistently outperform non-DC clustering algorithms for data integration tasks. However, we observed a significant correlation between the DC method and embedding approaches for rows, columns, and tables, highlighting that the suitable combination can enhance the efficiency of DC methods.
\end{abstract}



  \received{20 February 2007}
\received[revised]{12 March 2009}
\received[accepted]{5 June 2009}

\maketitle

\section{Introduction}
\label{sec:introduction}

Deep Learning (DL) is now a well-established machine learning paradigm that is effective in domains as diverse as image processing ~\cite{8917633}, natural language processing ~\cite{DBLP:journals/ijon/LauriolaLA22}, autonomous systems~\cite{DBLP:journals/tits/KuuttiBJBF21} and robotics~\cite{DBLP:journals/tce/BaiLLWL18a}. DL is also the subject of extensive investigation for data management tasks, including those relating to data cleaning and integration~\cite{DBLP:conf/edbt/Thirumuruganathan20}.

Deep Clustering (DC) is a sub-domain of DL in which deep neural networks are used to learn data representations in tandem with a clustering algorithm in an unsupervised manner. DC jointly optimizes the representation learning and clustering~\cite{DBLP:journals/access/MinGLZCL18}.  The importance of deep clustering is increasing due the need to deliver representation paradigms that can operate over increasingly heterogeneous and high-dimensional datasets ~\cite{DBLP:journals/corr/abs-1801-07648,DBLP:conf/aaai/MelloAM22}. DC also avoids the need for separate feature extraction, reduction and clustering ~\cite{DBLP:journals/corr/abs-1801-07648}.

DL has been applied successfully to a variety of data cleaning and integration problems, and several such problems involve clustering, so it seems timely to investigate the application of DC in data preparation. The approach in this paper is to empirically evaluate three deep clustering algorithms, comparing them to baselines that use  non-deep clustering techniques. For each of several problems, specifically {\it schema inference}, {\it entity resolution} and {\it domain discovery}, we: (i) define these tasks as clustering problems; (ii) identify several representations relevant to the tasks considering the type of data; (iii) compare the performance of three deep clustering algorithms against three representative non-deep clustering algorithms; and (iv) analyze the results in terms of overall quality and drill down to understand the behavioural properties of different techniques.

The contributions of this paper are as follows:
\begin{enumerate}
\item The identification of DC as a promising approach for data cleaning and integration tasks that stand to benefit from clustering.
\item The application of DC algorithms to {\it schema inference}, {\it entity resolution} and {\it domain discovery}, using vector representations for tables, rows and columns, respectively.
\item An empirical evaluation using third-party benchmarks that shows promising results; the selected DC algorithms consistently outperform non-DC clustering algorithms. The evaluation also provides insights on the configuration of DC algorithms for use with tabular data, and on the impact of different embeddings on clustering performance.
\end{enumerate}

The remainder of this paper is structured as follows. Section \ref{sec:related} outlines the development of work on DC. Section \ref{sec:dc} introduces key concepts in DC and describes the three algorithms used in the experiments. Section \ref{sec:experiment} describes the experimental methods that will be applied in the paper. Sections \ref{sec:si}, \ref{sec:er} and \ref{sec:dd} present the experiments on schema inference, entity resolution and domain discovery, respectively. Section \ref{sec:Conclusions and discussion} presents some conclusions and areas for further work.

\section{Related Work} \label{sec:related}
This section briefly reviews related work on DC and discusses the components of DC, such as representation learning and clustering. Furthermore, we review how both modules can be optimized in a single framework when applied to data integration problems.

Standard clustering (SC) methods have achieved significant success for various applications when the data is low dimensional and where there is the assumption that vectors in the latent space are well-shaped and, most of the time, linearly separable. However, SC methods struggle to effectively perform clustering without representation learning when the data is unstructured, high-dimensional, and heterogeneous ~\cite{DBLP:journals/corr/abs-2206-07579}. 
DC focuses on the joint optimization of high dimensional data representation in the latent space with suitability for clustering ~\cite{DBLP:journals/corr/abs-2206-07579}. DC enables interaction between (i) clustering and (ii) representation learning through joint optimization to improve both of them iteratively. 

Several proposals for clustering and representation learning architectures have been developed ~\cite{DBLP:journals/corr/abs-2206-07579}. The representation learning architectures take a raw high dimensional distance matrix as input and map it to a low dimensional latent space. In data management, the distance matrix is represented by distances between embeddings in the latent space.  

The most widely used representation learning architecture in deep clustering is Auto-encoder (AE) based unsupervised learning ~\cite{ DBLP:conf/ciarp/SongLHWT13, hinton2006reducing}. The encoder function $f_e$ encodes the input representation $x_i$ into a low dimensional representation \begin{math}h_i= f_e(x_i)=\frac{1}{1+e^{-(Wx_i+b_i)}}\end{math}, and the decoder function $f_d$ decodes $f_e$($x_i)$ into the reconstructed input ${\overline{x}}_i$= $f_d$($h_i)$.  $W$ and $b_i$ are the  weights and bias of neural networks. The optimization function of a simple AE architecture for $N$ samples can be defined as: 
\begin{math}
f_{min}={min \frac{1}{N}\sum^N_{i=1}{{\left\|x_i-\mathrm{\ }{\overline{x}}_i\right\|}^2}\ }.
\end{math}
Considering different applications, researchers proposed enhanced versions of AE, including Convolutional AE ~\cite{DBLP:conf/iccv/DizajiHDCH17} for image clustering, Variational AE ~\cite{DBLP:conf/aaai/XuSDT17} for text classification, Generative AE ~\cite{DBLP:journals/tnn/YeB22} for image reconstruction, and Adversarial AE ~\cite{DBLP:conf/nips/PidhorskyiAD18} to detect generative probabilistic novelty.

Feature distribution in the latent space is important; the learning efficiency depends on the features' distribution. In this context, subspace representation learning ~\cite{DBLP:conf/cvpr/CaiFGWZ022,DBLP:conf/nips/JiZLSR17,DBLP:conf/cvpr/ZhouHF18,DBLP:conf/cvpr/ZhangLYQZ0L19} has been used widely for clustering. In subspace representation learning, the latent spaces are divided into several subspaces to categorize instances, and two instances are associated with linear relationships in the same subspace.

Regarding the clustering architecture in DC, it takes the optimized low-dimensional latent representation as input and returns the clustering soft assignments. At this stage, the learned representations are evaluated to determine whether it is more cluster-friendly, for example, if two contextual instances are close to each other in latent space. Several clustering techniques have been used in deep clustering ~\cite{DBLP:journals/corr/abs-2206-07579}. The basic structure of the clustering component is to feed the $d$-dimensional representation using neural networks in the forward direction and reduce the dimensions to cluster number $K$. Then, a softmax layer can be used for the cluster assignment ~\cite{DBLP:journals/corr/abs-2206-07579}. 

To bridge the semantic gap between representation learning and clustering, relation-matching deep clustering techniques have been used ~\cite{DBLP:conf/cvpr/DangD0WH21, DBLP:conf/eccv/GansbekeVGPG20}, though such proposals are computationally expensive ~\cite{DBLP:journals/corr/abs-2206-07579}. A more advanced proposal includes graph-based architectures (e.g.,~\cite{DBLP:conf/www/Bo0SZL020,DBLP:conf/aaai/LiuTZLSYZ22,DBLP:conf/iccv/ZhongW0HDNL021} ) in which multiple distributions are generated and fed to graph neural networks to preserve the hidden relations between the latent and $K$-dimensional target distribution.

\section{Deep Clustering Concepts and Techniques}
\label{sec:dc}

The fundamental difference between SC and DC methods is that SC methods act on a static representation, and DC methods adapt and learn the representation and use it for clustering. Most SC methods follow a hard clustering mechanism that takes a distance matrix as input and returns the 1-dimensional discrete clustering labels~\cite{DBLP:journals/corr/abs-2206-07579}. It is hard to optimize a 1-dimensional discrete vector for a neural network. Instead, DC algorithms work on a soft clustering mechanism that takes a high dimensional distance matrix as input, learns the representation in a low dimensional latent space, and returns a K-dimensional continuous vector in the label space. The resulting K-dimensional continuous vector can be optimized for the final clustering 1-dimensional discrete vector ~\cite{DBLP:journals/corr/abs-2206-07579}.

The basic DC framework consists of three main components, i.e., (i) learning representation architecture, (ii) reconstruction loss, and (iii) clustering loss.

\begin{figure}[t]
  \centering\includegraphics[width=\linewidth,keepaspectratio]{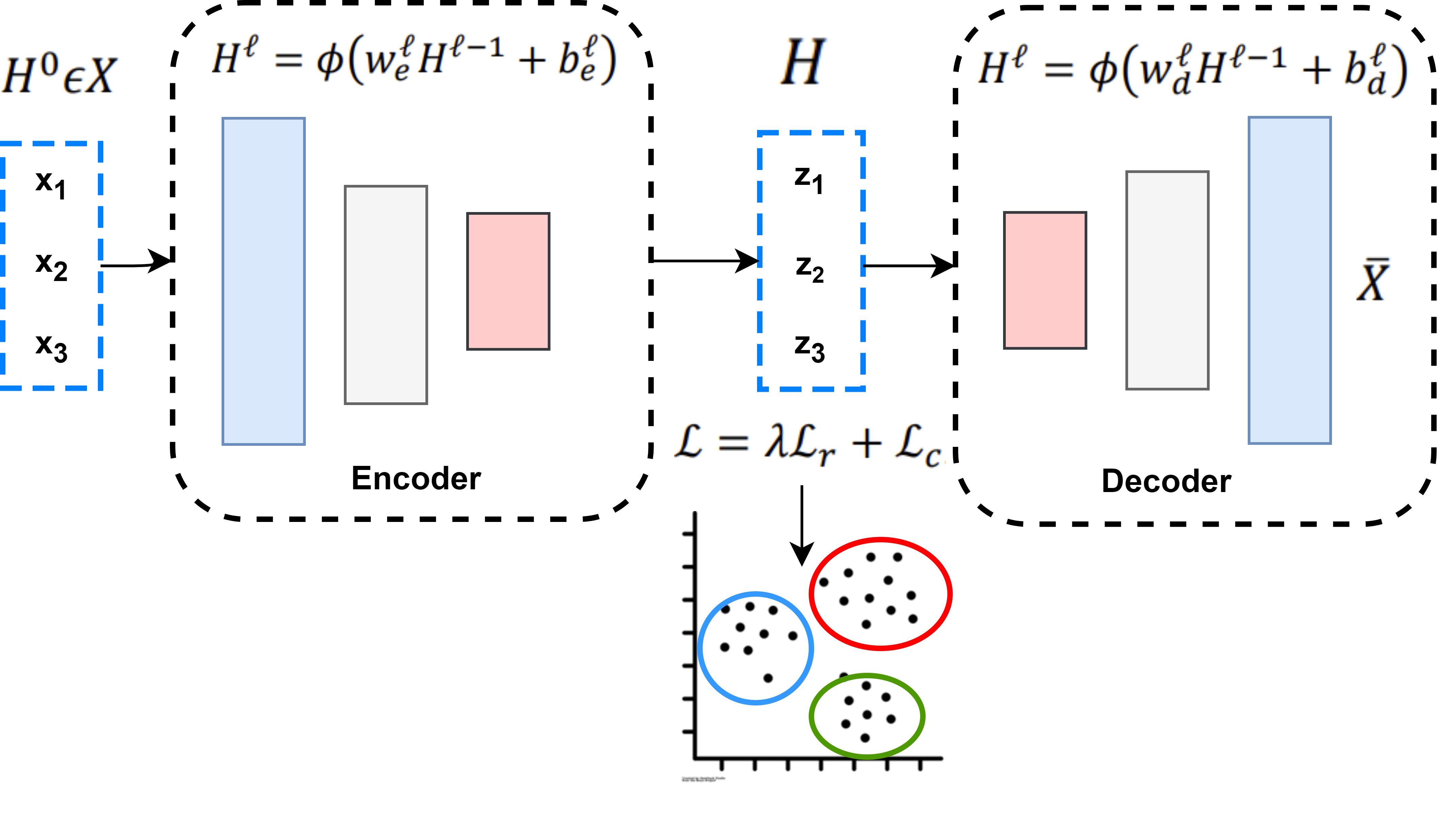}
  \caption{DC with basic AE architecture}
  \label{fig:framwork}
\end{figure}

Concerning (i), we adopt DC methods that use AE based and subspace representation learning architectures. A DC framework with a basic AE architecture is presented in Figure \ref{fig:framwork}. In an AE based architecture, the clustering is based on the lower dimensional representation produced by the encoder, with a loss function that trades off cluster quality with the ability to reconstruct the original representation. Through this approach, the latent space representation of the original input data should preserve the most suitable features for clustering.

Consider raw data $X\epsilon {\mathcal{R}}^{N\times d}$, where  ${\mathcal{R}}^{N\times d}$ belongs to a $d$- dimensional distance matrix  $\mathcal{R}$ with $N$ elements, $x_i$ is the $ith$ element in $X$.  Representation learning in AE initiates with the encoder part, the purpose of which is to encode $X$ into a low dimensional latent representation $H$. Let's suppose the AE consists of $L$ layers where $\ell $ is layer number, the initial representation learned from ${\mathcal{R}}^{N\times d}$ in encoder $H^{\ell }$ can be obtained as ~\cite{DBLP:conf/www/Bo0SZL020}:
\begin{equation}
\label{eq1}
H^{\ell }=\phi \left(w^{\ell }_eH^{\ell -1}+b^{\ell }_e\right),
\end{equation}
where $H^0\epsilon X$ and $\phi $ denotes the activation function, $w^{\ell }_e$ and $b^{\ell }_e$ represents the weight and bias of $\ell th$ layer. The decoder part decodes $H$ into reconstructed input $\overline{X}$ using the following equation ~\cite{DBLP:conf/www/Bo0SZL020}:
\begin{equation}
\label{eq2}
H^{\ell }=\phi \left(w^{\ell }_dH^{\ell -1}+b^{\ell }_d\right),
\end{equation}

where $H^0\epsilon \overline{X}$, $w^{\ell }_d$ and $b^{\ell }_d$ represents the weight and bias of $\ell th$ layer for decoder. The objective function used in AE architectures can be defined as:
\begin{equation}
\label{eq3}
\mathcal{L}=\lambda {\mathcal{L}}_r+{\mathcal{L}}_c,
\end{equation}
where ${\mathcal{L}}_r$ and ${\mathcal{L}}_c$ represent reconstruction and clustering loss, respectively. ${\mathcal{L}}_c$ is clustering module specific and each deep clustering proposal provides several other module specific losses that are combined with ${\mathcal{L}}_c$. The basic version of ${\mathcal{L}}_r$ can be defined as:

\begin{equation}
\label{eq4}
 f_{min}={min \frac{1}{N}\sum^N_{i=1}{{\left\|X-\mathrm{\ }{\overline{X}}\right\|}^2}\ }
\end{equation}

We adopted three recently proposed DC algorithms, SDCN ~\cite{DBLP:conf/www/Bo0SZL020}, EDESC ~\cite{DBLP:conf/cvpr/CaiFGWZ022} and SHGP ~\cite{DBLP:conf/nips/0002GW0XLH22}, to evaluate on data integration tasks. The selection of the DC algorithms is based on their implementation suitability and the flexibility of their proposed distance functions towards data integration tasks; many DC methods are purely designed for image and text clustering applications \cite{DBLP:conf/iccv/XingHXWXXWZS21,DBLP:conf/cvpr/Park0KKPHC21,DBLP:conf/iclr/TsaiLZ21} and are not obviously suitable for comparing rows, columns, and tables in the latent space. Another selection criterion is performance; we describe the top three performers on data integration tasks. 

The description of the DC algorithms used for the experimental evaluation is given below:

\begin{itemize}
\item \textbf{SDCN} ~\cite{DBLP:conf/www/Bo0SZL020} is based on two representation learning modules, a Graph Convolutional Network (GCN) and an AE, that work in parallel to learn structural and AE specific information. SDCN starts by constructing a K-Nearest Neighbor (KNN) graph from  $X$ and feeding it to the GCN model to learn the structural information. 
To learn AE-specific representations, SDCN uses a simple AE architecture. GCN-specific and AE-specific representations are combined through a delivery operator and dual self-supervised mechanism to perform soft clustering assignments from multiple representations. 


\item \textbf{EDESC} ~\cite{DBLP:conf/cvpr/CaiFGWZ022} is a deep subspace clustering method. Subspace representation learning involves mapping data points into low dimensional subspaces to separate each data point, similar to the early stage of subspace clustering ~\cite{DBLP:journals/corr/abs-2206-07579}.  Unlike SDCN, EDESC is not graph-based. 
Deep subspace clustering models are self-expressive and assume a linear combination between one data point and its other data points from the same subspace. The simplest self-expressiveness property can be denoted as $X=XC$, where $X$ represents the data matrix, and $C$ represents the self-expression coefficient matrix. The objective function for self-expression-based representation learning can be defined as ~\cite{DBLP:journals/corr/abs-2206-07579}:
\begin{equation}
\label{eq_EDESC}
{\mathop{\mathrm{min}}_{C} \ {\left\|C\right\|}_p\ }+\frac{\lambda }{2}{\ \left\|H-HC\right\|}^2_F\ \ \ \ \ \ s.t.\ \ \ \ \ \ \ \ diag\left(C\right)=0
\end{equation}
where ${\left\|.\right\|}_p$ shows matrix norm and $\lambda $ is weight controlling factor. $H$ is the representation learned by the network. EDESC takes a deep representation and learns the subspace bases in an iterative refining manner. The latent space representation is learned through the refined subspace bases outside the self-expressive framework. EDESC initializes a subspace D using K-means ~\cite{DBLP:conf/cvpr/CaiFGWZ022}.


\item \textbf{SHGP} ~\cite{DBLP:conf/nips/0002GW0XLH22} employs self-supervised learning on Heterogeneous Information Networks (HINs) and effectively uses a combination of attention-aggregation schemes using two modules, \textit{Att-LPA} and \textit{Att-HGNN}, that effectively improve each other to construct and learn object embeddings. The \textit{Att-LPA} module generates pseudo-labels, serving as a self-supervised signal to guide the learning process. \textit{Att-LPA} uses a structural clustering method (LPA), which assigns and iteratively refines the labels to objects. These pseudo-labels are then utilized as a guide to enhance the learning of object embeddings and attention coefficients in the \textit{Att-HGNN} module. \textit{Att-HGNN},  directed by the pseudo-labels, uses object features and attention coefficients to combine information from neighboring features and learns the embeddings effectively. 
For the clustering task, SHGP uses K-means on the embeddings produced by both modules.
\end{itemize}

SDCN and EDESC build on a pre-trained AE that learns a compressed representation of the optimized input embedding for reconstruction while ignoring the clustering task. AE can help reduce the input embedding's dimensionality, remove noise and redundancy, and capture relevant patterns and structure. Then, the learned representation passes to the original training part combined with clustering loss for further fine-tuning. It is helpful to evaluate the impact of learned representation on non-DC algorithms. In this context, we used a different AE version that employs the Birch and K-means algorithms to perform clustering not directly on the embedding but on the representation learned by AE. This can be interpreted as performing Birch and K-means on $H$ in Figure \ref{fig:framwork}.

\section{Experimental Setup}
\label{sec:experiment}

The hypothesis is that DC is expected to outperform SC as it builds on a latent space representation that can better integrate schema-level and instance-level representations. To evaluate the hypothesis, we included the following SC methods:

\begin{itemize}
\item \textbf{K-means} ~\cite{Hartigan_1979} initializes with a set of data points; in the context of the data integration problem, it initializes with distance vectors of either schema or instance-level data points in the vector space and assigns the data points to the clusters with the nearest centroid. It repeatedly iterates to optimize the cluster centers. K-means minimizes the clustering loss with a squared Euclidean distance function. K-means requires the value of K in advance to predict the clusters. 


\item \textbf{Birch} ~\cite{DBLP:conf/sigmod/ZhangRL96} is from a hierarchical clustering family designed for tackling large databases, especially involving data with noise and outliers. 
Birch is supervised in terms of number of clusters $K$.  
Birch provides a hierarchical clustering structure, which can help understand the data structure and provide more interpretable results. 

\item \textbf{DBSCAN} ~\cite{DBLP:conf/kdd/EsterKSX96} is a density-based spatial clustering algorithm primarily designed for large databases with noise. DBSCAN identifies clusters based on the density of the data; it has two main parameters: the radius $ \varepsilon $, which defines the area in the neighboring points, and \textit{MinPts}, which represents a minimum number of points required to declare the area dense enough to form a cluster. Unlike K-means and Birch, DBSCAN is suitable for identifying clusters with irregular shapes and does not require specifying the number of clusters $K$ in advance. DBSCAN is sensitive to its parameters $ \varepsilon $ and \textit{MinPts} and tends to provide poor clustering without optimizing its parameters. In our experiments, we used a commonly used heuristic method called the elbow method ~\cite{DBLP:journals/tods/SchubertSEKX17} to identify $ \varepsilon $. In the elbow method, the distances of data points to their nearest neighbors are calculated and plotted on a graph, and the "elbow" point is the best $ \varepsilon $-value where the curve intersects. The \textit{MinPts} are set to the total number of clusters $K$ when the rule of thumb (\textit{MinPts} = 2$\times$dim, where dim= the data dimensions) does not work.

\end{itemize}

Setting the number of clusters $K$ in advance gives the SC methods an (in a sense unfair) advantage as they are given the (unknown) Gound Truth (GT) value for $K$. In contrast, DC methods only take $K$ to initialize the centers of the clusters for pre-training. Subsequently, the DC methods work out $K$ without taking a GT value in training, and in practice, it may not be possible to establish the correct $K$. As such, the DC methods are more flexible in their ability to automatically estimate the number of clusters without the need for prior knowledge of $K$.

\begin{figure*}[!htbp]
  \centering
  \includegraphics[width=\linewidth,keepaspectratio]{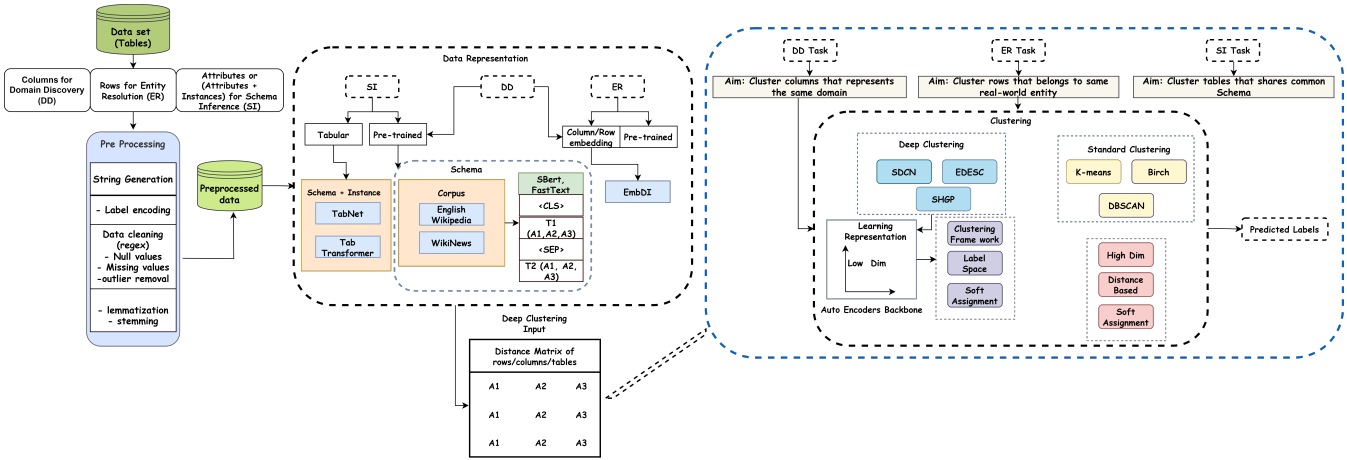}
  \caption{Overview of the experimental framework. Deep learning input will be considered as a distance matrix for each data integration problem}
  \label{fig:Expframwork}
\end{figure*}

We used the scikit-learn implementations ~\cite{scikit-learn} of the SC algorithms.  A detailed overview of the experimental framework is presented in Figure \ref{fig:Expframwork}, which consists of three main phases from left to right. Firstly, the raw data is preprocessed to remove high-level syntactic errors. In the second phase, the preprocessed data is fed to the embedding module to generate dense representations. Lastly, dense representations are further enhanced in the clustering module and final clustering assignments are produced.

\subsection{Evaluation Metrics}
\label{Eva_metrics}
We employ two widely used standard clustering evaluation metrics, Accuracy (ACC) ~\cite{DBLP:journals/tip/YangXNYZ10} and Adjusted Rand Score (ARI) ~\cite{DBLP:conf/iccv/WuLWQLLZ19}. 

 ARI can be defined as ~\cite{DBLP:conf/iccv/WuLWQLLZ19}: Assume we are given a set $S$ of $n$ elements and two clustering sets of these elements consists of $r$ and $s$ groups represented as $X=\left\{X_1,\ X_2,\dots ,X_r\right\}$ and $\{Y=Y_1,\ Y_2,\dots ,Y_s$$\mathrm{\}}$ in a contingency Table $\left[t_{ij}\right]$ of overlaps between $X$ and $Y$. Each element in ${[t}_{ij}]$ shows the count of common objects between $X_i$ and $Y_j$. ARI can be defined from ${[t}_{ij}]$:

 \begin{equation}
\label{eq5}
ARI=\frac{{\sum_{ij}{\left( \begin{array}{c}
t_{ij} \\ 
2 \end{array}
\right)}-\left[\sum_i{\left( \begin{array}{c}
a_i \\ 
2 \end{array}
\right)}\sum_j{\left( \begin{array}{c}
b_i \\ 
2 \end{array}
\right)}\right]}/{\left( \begin{array}{c}
n \\ 
2 \end{array}
\right)}}{{\frac{1}{2}\left[\sum_i{\left( \begin{array}{c}
a_i \\ 
2 \end{array}
\right)}+\sum_j{\left( \begin{array}{c}
b_i \\ 
2 \end{array}
\right)}\right]-\left[\sum_i{\left( \begin{array}{c}
a_i \\ 
2 \end{array}
\right)}\sum_j{\left( \begin{array}{c}
b_i \\ 
2 \end{array}
\right)}\right]}/{\left( \begin{array}{c}
n \\ 
2 \end{array}
\right)}} 
\end{equation}

ARI determines the similarity between two clustering results; usually, one is GT labels, and the second corresponds to the labels returned by the clustering algorithm. Generally, the value of ARI lies between 0 and 1. An ARI value closer to 1 represents a strong match between predicted and GT clusters.

The clustering ACC for $N$ samples, with cluster id $\mathrm{R}\in c_i$ and GT id ${T\in gt}_i$ can be defined as:
\begin{equation}
\label{eq6}
ACC\left(R,T\right)=\frac{\sum^N_{i=1}{\delta ({gt}_i,~~~map(c_i))}}{N}
\end{equation}
\begin{equation}
\label{eq7}
\delta \left({gt}_i,~map\left(c_i\right)\right)=\left\{ \begin{array}{c}
1,~~~~~~~~~~~if~{gt}_i=map\left(c_i\right) \\ 
0,\qquad otherwise~~~~ \end{array}
\right.
\end{equation}
Function $map()$ gives the best permutation mapping between predicted and GT labels through the Hungarian Algorithm ~\cite{DBLP:journals/tkde/CaiHH05}. ACC maps the predicted labels into ground labels since cluster ids in the prediction are randomly generated and dissimilar to those assigned to GT labels.

\subsection{Hyper-parameter setting}

 Network learning benefits from hyper-parameters optimized for the particular task. Since deep clustering algorithms are heavily used for image processing tasks, optimizing hyper-parameters for data integration tasks is necessary. In this context, we have three basic parameters of SDCN, EDESC and SHGP:

\begin{itemize}
\item \textbf{The number of layers} are important because they determine the network's ability to learn unsupervised feature representations. Since we have pre-defined embeddings as AE inputs, rows, columns or tables with similar meanings are positioned closer together. An AE with fewer layers can efficiently compress and reconstruct this underlying structure. We fixed \textit{number of layers = 2} in all experiments of SDCN and EDESC after experimenting with different values. However, SHGP uses \textit{Att-LPA} and \textit{Att-HGNN} with several layers from graph neural networks, so we used the default number of layers based on the SHGP paper, i.e., two \textit{Att-HGNN } encoder layers and several hidden layers in the set ${64, 128, 256, 512}$.

\item \textbf{Layer size} (refers to the number of neurons in each layer) provides the capacity and ability of AE to learn complex patterns in the data. In order to maintain a high-dimensional hidden representation, we fixed \textit{layer size = 1000} for all experiments of DC methods after experimenting with different values. This suggests that the complexity of row, column, or table embeddings requires a larger hidden layer size to retain more semantic information.
 
 \item \textbf{latent space size $z$}: Originally, SDCN and EDESC used $z= 10$, but it is too small in data integration tasks to capture the complexity of the row, column, or table embeddings, leading to significant information loss. Considering this, after systematically experimenting with different values, we fixed $\textit{z=100}$ for SDCN and AE, and $z=a$ for EDESC where the shape of $a= (n\_clusters \times d)$, and $d$ represents the dimension of the subspace. For SHGP, the size of the learned representation depended on the size of label space and appeared optimal after experimental evaluation as $z=s$ where $s = size of label space$.
 
\item \textbf{Training Epochs}: We used the silhouette coefficient ~\cite{rousseeuw1987silhouettes} on the learned representation with predicted clusters to choose where to stop training. We pre-train SDCN and EDESC for 30 epochs except for entity resolution (100 epochs), which requires more pre-training due to the large numbers of clusters. We decide the number of training epochs based on the best silhouette score. Since SHGP uses K-means for clustering the learned embedding, the embeddings obtained on 50 and above epochs do not significantly impact K-means clustering. We used fixed 50 epochs for representation learning by \textit{Att-LPA} and \textit{Att-HGNN}.
 \end{itemize}

We use AE (described in Section \ref{sec:dc}) with Birch for the entity resolution and domain discovery experiment instead of SDCN. The learned features without clustering loss from the AE step were more effective at capturing the underlying structure of the data than fine-tuning features along with clustering loss. In order to decide whether to use only AE for training or to continue for SDCN, we use the silhouette score. If the silhouette score converges during training with SDCN, we use SDCN; otherwise, we retain AE and cluster using Birch. The hyperparameter settings for the experiments, along with the source code, are given\footnote{\url{https://github.com/hafizrauf/dc_data-integration}}.

\section{Schema Inference}
\label{sec:si}

Schema inference proposes a schema that makes recurring structural features in data explicit. Schema inference may be applied to extensional data (e.g., inferring a JSON schema from several JSON documents ~\cite{DBLP:journals/vldb/BaaziziCGS19}) or to intensional data (e.g., inferring a schema that summarises a complex relational database~\cite{DBLP:conf/vldb/YuJ06a}). Schema inference has been a topic of ongoing investigation for different data models for some time, and several surveys have been produced~\cite{DBLP:conf/dateso/KlimekN10,DBLP:journals/vldb/CebiricGKKMTZ19,Kellou-Menouer-22}. It is common for schema inference to build on clustering, to identify candidate types/classes in the data, so clustering is an important enabling technology for schema inference. We consider schema inference as a clustering problem, where the task is: for a given set of datasets \textbf{$D=\{d_1,d_2,\ d_3\dots d_n\}$} identify every subset $D_s\subseteq D$ that can share a common schema using clustering.

Pre-trained embeddings have been used widely for data integration tasks ~\cite{DBLP:journals/pvldb/EbraheemTJOT18}. They fall into two specific categories: sentence-based and word based. Sentence-based embeddings directly map a sequence of tokens into a single dense vector. In contrast, word-based embeddings encode each token separately, and then perform an aggregation function to derive a single vector. Pre-trained embeddings are trained on large corpora and tend to have broader vocabulary coverage. Considering this, we choose two pre-trained embeddings (one word-based, FastText~\cite{DBLP:conf/lrec/GraveBGJM18}, and one sentence-based, SBERT ~\cite{Reimers_2019}) to perform schema inference with schema-level evidence. When carrying out schema inference, the schema-level information includes only table headers; each table is represented by a string (combination of attribute names). Nevertheless, pre-trained embeddings have disadvantages when tackling large databases, especially with instance level data, e.g., where there is specialised vocabulary, or numerical data distributions ~\cite{DBLP:conf/sigmod/CappuzzoPT20}.  To produce embeddings with Schema$+$instance-level data, tabular transformers have been a mainstream option in the deep learning community ~\cite{DBLP:journals/pvldb/BadaroP22}. Several tabular transformers are proposed in the literature to handle noisy and incomplete data (e.g., ~\cite{DBLP:journals/corr/abs-2012-06678,DBLP:conf/aaai/ArikP21,DBLP:journals/corr/abs-2106-01342,DBLP:conf/nips/GorishniyRKB21}).

For evaluation, we use the T2D Entity-Level Gold standard (T2Dv1)~\cite{DBLP:conf/edbt/RitzeB17} web table dataset and the Table Union Search (TUS) benchmark \cite{DBLP:journals/pvldb/NargesianZPM18} that identifies tables that are unionable. In T2Dv1, we rejected all tables that included languages other than English. We also excluded all DBpedia {\it Thing} tables to avoid significant data imbalance, as it is mapped to more than 50\% of data tables.  In the TUS benchmark, we aim to determine which tables from a set can be unioned together, and we set the criteria that two tables are unionable if at least 40\% of their corresponding attributes are unionable. First, we identify the unionable columns between table pairs. We then calculate the percentage of these unionable columns relative to the total columns of each table in a pair. We filter pairs using a threshold of 40\% unionable columns. These filtered pairs form a network's basis, with tables as nodes and shared columns as edges. We cluster these tables using the Louvain community detection algorithm \cite{blondel2008fast}. Each detected community, representing a group of unionable tables, and is assigned a unique GT label. We excluded all single-table communities. Further properties of the web tables data and TUS benchmark are given in Table \ref{tab:dataset_description}. The criteria for choosing a tabular transformer are based on the nature of the datasets. Web tables are noisy; for example, a table with {attribute: symbol} and values {'aa', 'axp'} is problematic for pre-trained embeddings as {'aa', 'axp'} are not present in the pre-trained vocabulary, and most of the cases will be treated as unknown tokens. However, training on a local vocabulary can overcome this issue. Another significant issue is incomplete columns or rows. To handle these issues, we evaluated six tabular transformers, including Tabnet ~\cite{DBLP:conf/aaai/ArikP21}, TabTransformer ~\cite{DBLP:journals/corr/abs-2012-06678}, SAINT ~\cite{DBLP:journals/corr/abs-2106-01342}, FT-Transformer ~\cite{DBLP:conf/nips/GorishniyRKB21}, TabFastFormer ~\cite{DBLP:conf/emnlp/KimH20}, and TabPerceiver ~\cite{DBLP:conf/icml/JaegleGBVZC21}, on web tables and TUS. We retained the two best performers, Tabnet and TabTransformer for comparison.  TabTransformer ~\cite{DBLP:journals/corr/abs-2012-06678} has been found to be robust with missing table values and noisy data. TabTransformer is based on Transformers ~\cite{DBLP:conf/nips/VaswaniSPUJGKP17} with several multi-head attention layers to contextually embed categorical columns. 
Tabnet ~\cite{DBLP:conf/aaai/ArikP21} is based on row-wise feature selection and is more suitable for raw data without pre-processing. Tabnet uses sequential attention to choose categorical and numerical features at each decision step. 

\begin{table*}[!htbp]
  \caption{Dataset properties for schema inference, entity resolution and domain discovery}
  \label{tab:dataset_description}
  \begin{tabular}{ccccccc}
	\toprule  
Properties & \multicolumn{2}{c}{\textbf{Schema Inference}} & \multicolumn{2}{c}{\textbf{Entity Resolution}} & \multicolumn{2}{c}{\textbf{Domain Discovery}} \\  
 & web tables & TUS & Music Brainz 2K & Geographic Settlements & Camera & Monitor \\  
 \midrule
\textbf{Sources} & N/A & N/A & 5 & 4 & 24 & 26 \\  
\textbf{Number of Instances} & 429 & 4248 & 2002 & 3021 & 19036 & 34481 \\  
\textbf{GT clusters} & 26 & 37 & 684 & 786 & 56 & 81 \\  
\bottomrule
\end{tabular}
\end{table*}

\subsection{Data dimensionality for tabular transformers}

To produce a distance matrix ($X_i$), each tabular encoding method produces a different dimension size $d$. In our experiments, we use the standard values of $d$ for FastText and SBERT as 300 and 768, respectively. Tabnet and TabTransformer process each input feature individually, whether row or column and apply a series of transformations. Each table's categorical and continuous features have different cardinalities affecting the size of output embedding $d$ for each table in Tabnet and TabTransformer. To normalize $d$ for instance-level data, we selected the maximum feature size occurrence and performed linear interpolation to fill the empty values. However, for TabTransformer, the last column of the distance matrix needs the preceding value to interpolate, which makes the size of $d$ as $max($d$ -1)$, where $max(d)$ denotes the maximum number of dimensions observed for any table. The obtained values of $d$ for Tabnet and TabTransformer are 693 and 208 for web tables 1365 and 352 for TUS data, respectively. 

\subsection{Results and Discussion}

\begin{figure}[!t]
     \begin{subfigure}[b]{0.23\textwidth}
         \centering
         \includegraphics[width=1.2\linewidth]{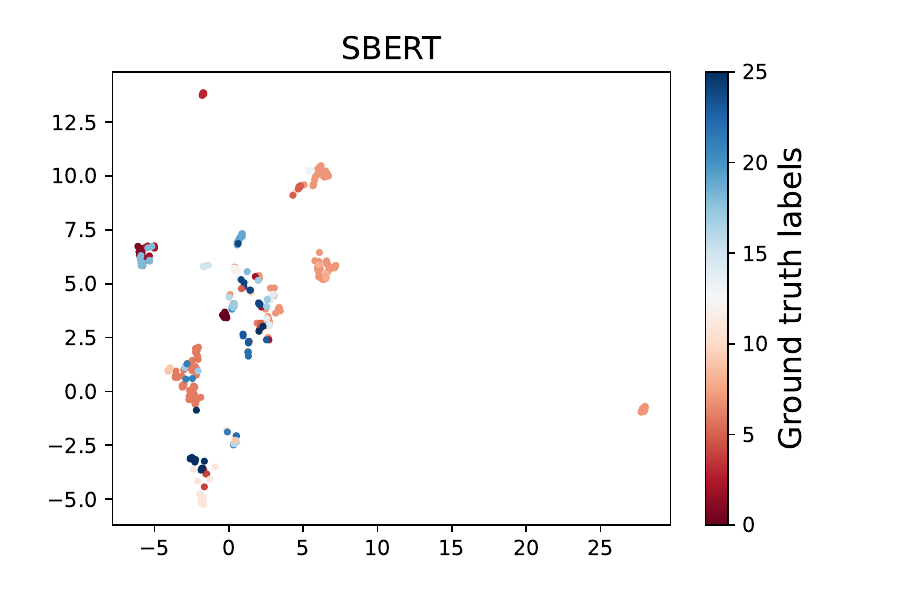}
         \caption{Schema-level}
         \label{fig:si:1a}
     \end{subfigure}
     \begin{subfigure}[b]{0.23\textwidth}
         \centering
         \includegraphics[width=1.2\linewidth]{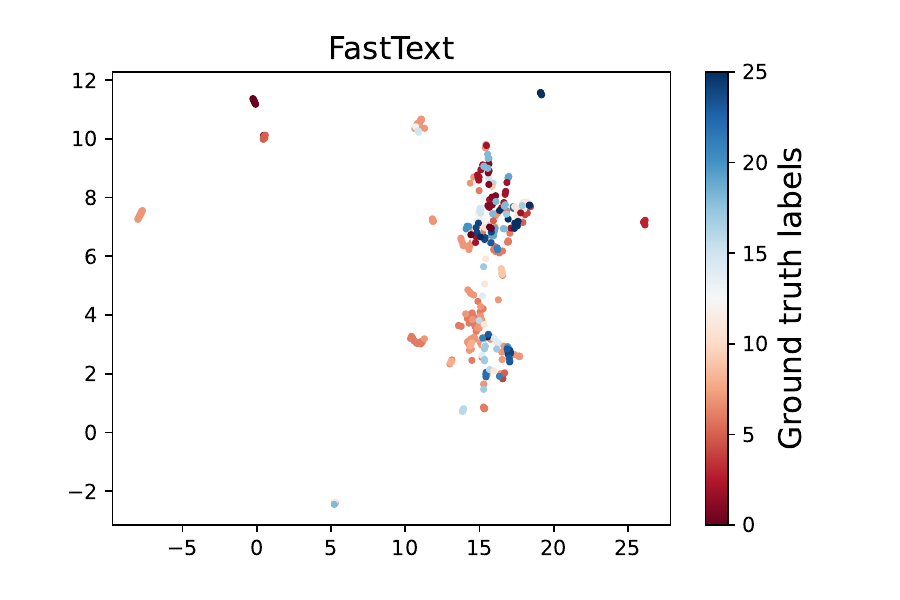}
         \caption{Schema-level}
         \label{fig:si:1b}
     \end{subfigure}
     \begin{subfigure}[b]{0.23\textwidth}
         \centering
         \includegraphics[width=1.2\linewidth]{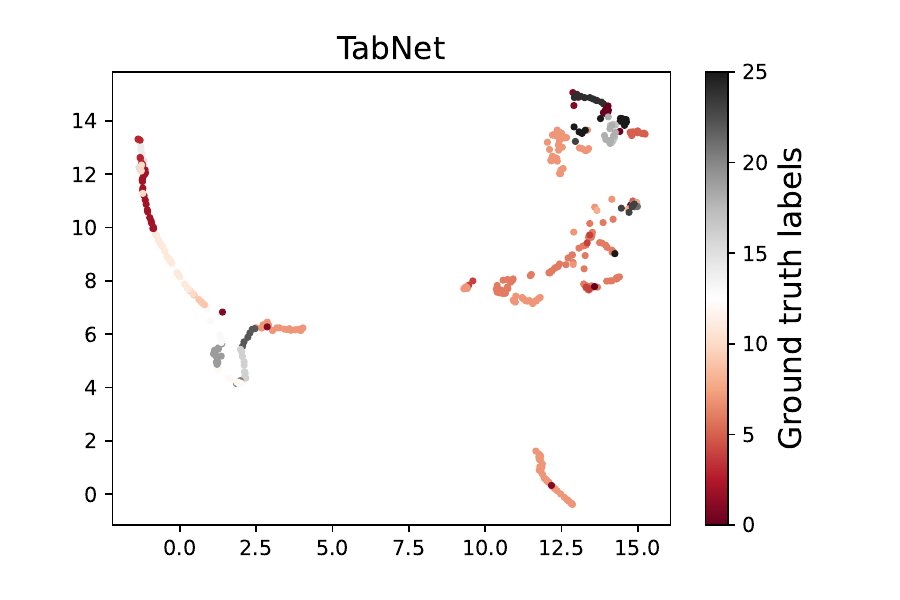}
         \caption{Schema$+$Instance-level}
         \label{fig:si:1c}
     \end{subfigure}
     \begin{subfigure}[b]{0.23\textwidth}
         \centering
         \includegraphics[width=1.2\linewidth]{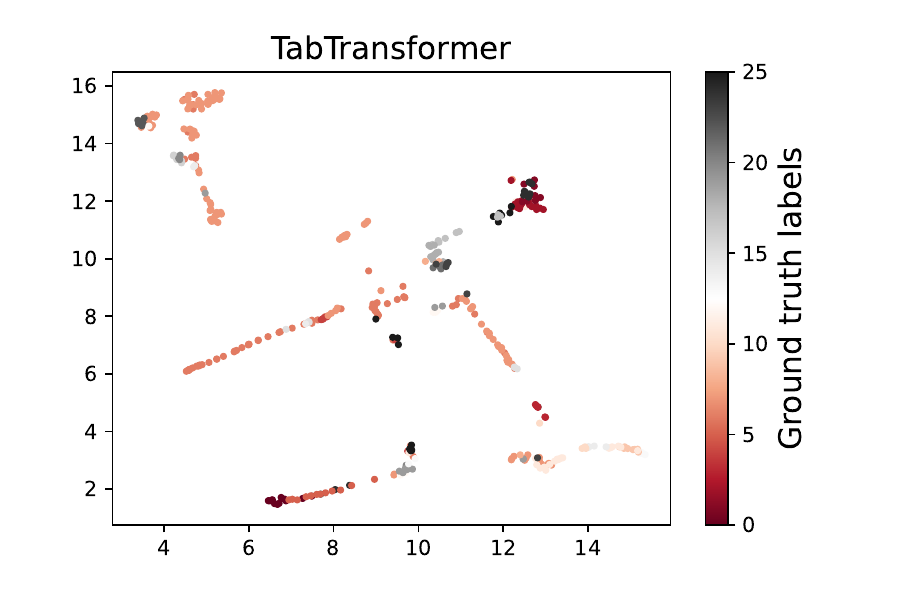}
         \caption{Schema$+$Instance-level}
         \label{fig:si:1d}
     \end{subfigure}
     \caption{Umap representation of pre-trained sentence and tabular based encodings on web tables data}
     \label{fig:si:1}
\end{figure}

For all experimental results, the bold and underlined values in the tables indicate the best and the second-best results considering the corresponding embedding methods, respectively. Table \ref{tab:siresult:1} presents the clustering results for schema inference using only schema-level data.  The following can be observed:

(i) \textbf{The selected table representation significantly impacts performance, with SBERT significantly outperforming FastText in most cases for all clustering algorithms on all datasets}. SDCN achieved 0.38 higher ARI with SBERT than FastText, where Birch and K-means with SBERT outperformed Fast Text by 0.34 and 0.17 in the ARI on web tables data. Similarly, EDESC and SHGP with SBERT are superior by 0.23 and 0.13 ARI than FastText on TUS data.
Figures \ref{fig:si:1a} and \ref{fig:si:1b} confirm that the separability of data points for SBERT is more robust than for FastText, in which data points are compact in the latent space, which is unsuitable for clustering. 
(ii) \textbf{The DC algorithms outperform the SC algorithms in most cases, with the largest differences being for SBERT-supported models}. EDESC with SBERT obtained higher ARI scores (0.15, 0.18, and 0.66 ) on TUS data compared to K-means, DBSCAN, and Birch, respectively. K-means, which considers convex and isotropic clusters, obtained a modest ACC difference of 0.09, signifying its likely struggles with dense data. Birch did better, and achieved an ARI of 0.33, but still lower than SDCN (0.13 ARI) and EDESC (0.08 ARI) on web tables data. DBSCAN predicted one cluster and zero ARI, which shows its lower ability with varying data densities. (iii) \textbf{SDCN predicted fewer clusters than other competitors while maintaining superior performance.} SDCN with SBERT predicted 16 clusters, diverging from the 26 predicted by EDESC and SHGP, matching the GT count. However, SDCN's superior performance compared to EDESC and SHGP indicates its AE's ability to prioritize clusters' internal coherence and quality over count. In contrast, EDESC and SHGP failed to maintain the inter-cluster separability, even when meeting the GT count.


\begin{table*}[!htbp]
  \caption{Schema Inference: Schema-level clustering results DC (SDCN, EDESC and SHGP) vs SC (K-means, Birch and DBSCAN) using pre-trained embeddings on web tables and TUS datasets.}
  \label{tab:siresult:1}
 \begin{tabular}{cccccccccccccc} 
	\toprule 
 &  & \multicolumn{2}{c}{SDCN} & \multicolumn{2}{c}{SHGP} & \multicolumn{2}{c}{EDESC} & \multicolumn{2}{c}{K-means} & \multicolumn{2}{c}{DBSCAN} & \multicolumn{2}{c}{Birch} \\ 
  \midrule 
\textbf{Dataset} & \textbf{Metric} & SBERT & FastText & SBERT & FastText & SBERT & FastText & SBERT & FastText & SBERT & FastText & SBERT & FastText \\  
  & $K$ & 16 & 19 & 26 & 26 & 26 & 26 & 26 & 26 & 1 & 1 & 26 & 26 \\  
 web & ARI & \textbf{0.46} & 0.08 & 0.10 & 0.05 & \underline{0.41} & \textbf{0.14} & 0.27 & \underline{0.10} & 0.0 & -0.018 & 0.33 & -0.01 \\  
 tables & ACC & \textbf{0.58} & 0.27 & 0.32 & 0.27 & \underline{0.55} & \textbf{0.35} & 0.45 & \underline{0.31} & 0.29 & 0.24 & 0.49 & 0.28 \\  
  & $K$ & 37 & 33 & 37 & 37 & 37 & 36 & 37 & 37 & 18 & 4 & 12 & 1 \\  
 TUS & ARI & \underline{0.74} & \textbf{0.70} & 0.65	& 0.53 & \textbf{0.88} & \underline{0.65} & 0.73 & 0.63 & 0.70 & 0.05 & 0.22 & 0.0 \\  
 & ACC & \underline{0.79} & \textbf{0.74} & 0.73& 0.63& \textbf{0.87}& \underline{0.73} & \underline{0.79} & 0.69 & 0.78 & 0.29 & 0.40 & 0.20 \\  
 \bottomrule
\end{tabular}
\end{table*}

Table \ref{tab:siresult:2} presents the clustering results for schema inference using both schema and instance level data.  The following can be observed: (i) \textbf{SDCN is significantly more compatible with Tabnet than TabTransformer on all datasets}. SDCN with Tabnet obtained an ARI score 0.19 and 0.13 higher than TabTransformer on web tables and TUS data, respectively. This superior performance exhibits Tabnet's ability to  select prominent features at each decision step, thus improving SDCN's feature extraction process with more detailed data understanding. Figures \ref{fig:si:1c} and \ref{fig:si:1d} represent no significant latent space difference in terms of data points' relative positions. This low visual difference entirely depends on the Umap projections, which can lead to the loss of information in relationships among the tables. Furthermore, it shows that web table data does not have a clear cluster structure, making it difficult to discover meaningful patterns. Adding instance-level evidence with tabular embedding failed to show its suitability for clustering compared with schema-level evidence with SBERT for both datasets.  (ii) \textbf{Changing the embeddings does not affect the overall performance trend for the DC method when we consider Schema$+$Instance-level data}. We observe that DC methods outperformed SC methods with both Tabnet (SDCN obtained 0.46, 0.45, 0.46 higher ARI compared to K-means, DBSCAN, and Birch, respectively) and TabTransformer (SDCN obtained 0.24 higher ARI compared to K-means, DBSCAN, and Birch) on web tables data. Like schema-level in Schema$+$Instance-level, DBSCAN's repeated inability to effectively differentiate clusters within the dense data representation resulted in a single cluster for Tabnet and TabTransformer on web tables data. 
(iii) \textbf{The provision of K does not significantly impact the clustering algorithms' overall performance. }  For example, Birch with TabTrasnformer may have been expected to outperform EDESC due to the prescription of a fixed number of clusters (26), but EDESC outperformed Birch by 0.07 ARI even though it only produced 14 clusters. Similar behavior can be seen when using Tabnet with EDESC, which produced 12 clusters compared to 26 GT clusters and achieved 0.09 higher ARI than K-means, which generated 26 clusters on web tables data. On the other hand, DBCSCAN produced 3 clusters on TUS data but failed to produce convincing clusters, compared EDESC with an exact number of precdited clusters 37 as GT. (iv) \textbf{Tabnet and TabTransformer treat all attributes as being equally important.}  As a result, even when two tables share a subject attribute, they may be clustered separately because their other attributes are different. A {\it subject attribute} identifies the artifact that the table is about. For example in web tables data, tables T1 and T2, are clustered separately where they have a common subject column \textit{Country} and other columns \textit{(T1.Total population in 2004 (million), T1.Annual population growth rate (\%), T1.Population density (persons per square km.), T1.Average number of persons per household)} and \textit{(T2.rank, T2.population, T2.date of information)}.

\begin{table*}[!htbp]
  \caption{Schema Inference: Schema$+$Instance-level clustering results DC (SDCN, EDESC and SHGP) vs SC (K-means, Birch and DBSCAN) using tabular embeddings on web tables and TUS datasets. TT and TN refer to TabTransformer and Tabnet, respectively.}
  \label{tab:siresult:2}
  \begin{tabular}{cccccccccccccc} 
	\toprule 
 &  & \multicolumn{2}{c}{SDCN} & \multicolumn{2}{c}{SHGP} & \multicolumn{2}{c}{EDESC} & \multicolumn{2}{c}{K-means} & \multicolumn{2}{c}{DBSCAN} & \multicolumn{2}{c}{Birch} \\ 
 \midrule 
\textbf{Dataset} & \textbf{Metric} & TT & TN & TT & TN & TT & TN & TT & TN & TT & TN & TT & TN \\  
 & $K$ & 26 & 26 & 26 & 26 & 14 & 12 & 26 & 26 & 1 & 1 & 26 & 26 \\  
 web tables  & ARI & \textbf{0.26} & \textbf{0.45} & 0.02 & -0.019 & \underline{0.09} &  \underline{0.08} & 0.02 & -0.013 & 0.023 & -0.007 & 0.02 & -0.013 \\  
 & ACC & \textbf{0.42} & \textbf{0.55} & 0.29 & 0.25 & \underline{0.31} & \underline{ 0.31} & 0.29 & 0.27 & 0.29 & 0.26 & 0.28 & 0.27 \\  
  & $K$ & 37 & 37 & 37 & 37 & 37 & 37 & 37 & 37 & 3 & 3 & 37 & 37 \\  
TUS & ARI & \textbf{0.29} & \textbf{0.34} & 0.06 & 0.06 &  \underline{0.24} & 0.25 &  0.21 & 0.25 & 0.02 & 0.03 & 0.18 &  \underline{0.26} \\  
 & ACC & \textbf{0.44} & \textbf{0.45} & 0.21 & 0.21 & \underline{0.40} &  \underline{0.38} &  \underline{0.38} & 0.38 & 0.26 & 0.26 & 0.35 &  \underline{0.38} \\ 
 \bottomrule 
\end{tabular}
\end{table*}

In terms of relative performance between Tables \ref{tab:siresult:1} and \ref{tab:siresult:2}, empirical results for the web tables dataset show that: \textbf{schema-level evidence is more suitable for DC and SC, and adding instance-level evidence leads to poorer performance}.  This is because the actual instances tend to have low overlap even when their tables are clustered together in the GT. For example, SDCN with Tabnet failed to cluster tables T3 and T4, which belong to the class {\it Film} because of the same schema and different instances, e.g., (\textit{T3.fansrank: 101, T3.title: treasure sierra madre, T3.year: 1948, T3.director: john huston, T3.overallrank: 92)} and \textit{(T4.fansrank: 442, T4.title: game, T4.year: 1997, T4.director: david fincher, T4.overallrank: 1491).} 

Overall, considering the observations and evidence from Tables \ref{tab:siresult:1} and \ref{tab:siresult:2}, we can conclude that DC outperforms SC, regardless of the selected embedding strategy.

\section{Entity Resolution}
\label{sec:er}

Entity resolution is the well-studied process of identifying where two or more records in a dataset represent the same real world object~\cite{DBLP:journals/csur/ChristophidesEP21,DBLP:journals/tkde/ElmagarmidIV07}. Entity resolution thus takes place at the instance level. Most entity resolution proposals focus on pairwise similarity between records. However, the transitive closure of the pairwise similarity relationship may not lead to suitable clusters, and as a result some entity resolution proposals include a clustering step (e.g., \cite{DBLP:journals/datamine/CostaMO10,DBLP:journals/pvldb/HassanzadehCML09}). We note that deep learning has been applied with positive results to entity resolution (e.g., \cite{DBLP:journals/pvldb/EbraheemTJOT18,DBLP:journals/pvldb/Thirumuruganathan21}), but that the need for training data is a barrier to adoption~\cite{DBLP:conf/icde/BogatuPDDF21}. Deep clustering can potentially provide some of the benefits of deep representation learning in an unsupervised setting. The task of entity resolutioon with clustering can be defined as:  given a set of records $R=\{r_1,r_2,\ r_3\dots r_n\}$ identify every subset $R_s\subseteq R$ that refers to the same real world entity using clustering.

For the entity resolution task, we employed the MusicBrainz \cite{DBLP:conf/adbis/SaeediPR17} and Geographic Settlements datasets \cite{DBLP:conf/adbis/SaeediPR17}. MusicBrainz contains continuously updated song data from five sources and includes duplicates for 50\% of the original records, whereas Geographic Settlements contains geographical real-world entities from four data sources.  

\begin{figure}[t]
    \centering
    \begin{subfigure}[b]{0.48\linewidth}
        \centering
        \includegraphics[width=\linewidth]{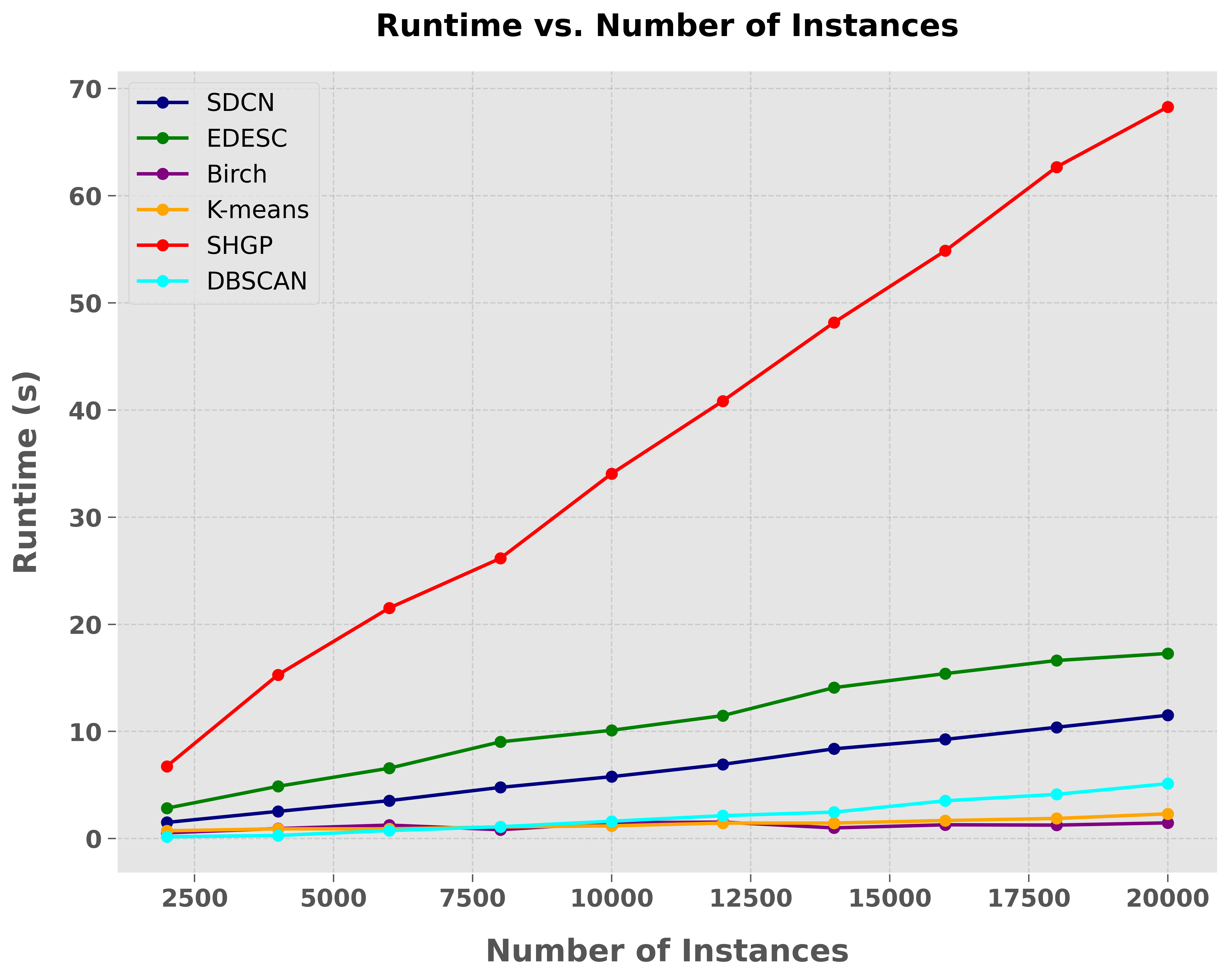}
        \caption{Runtime vs. Number of Instances.}
        \label{fig:er:runtime-inst}
    \end{subfigure}
    \hfill
    \begin{subfigure}[b]{0.48\linewidth}
        \centering
        \includegraphics[width=\linewidth]{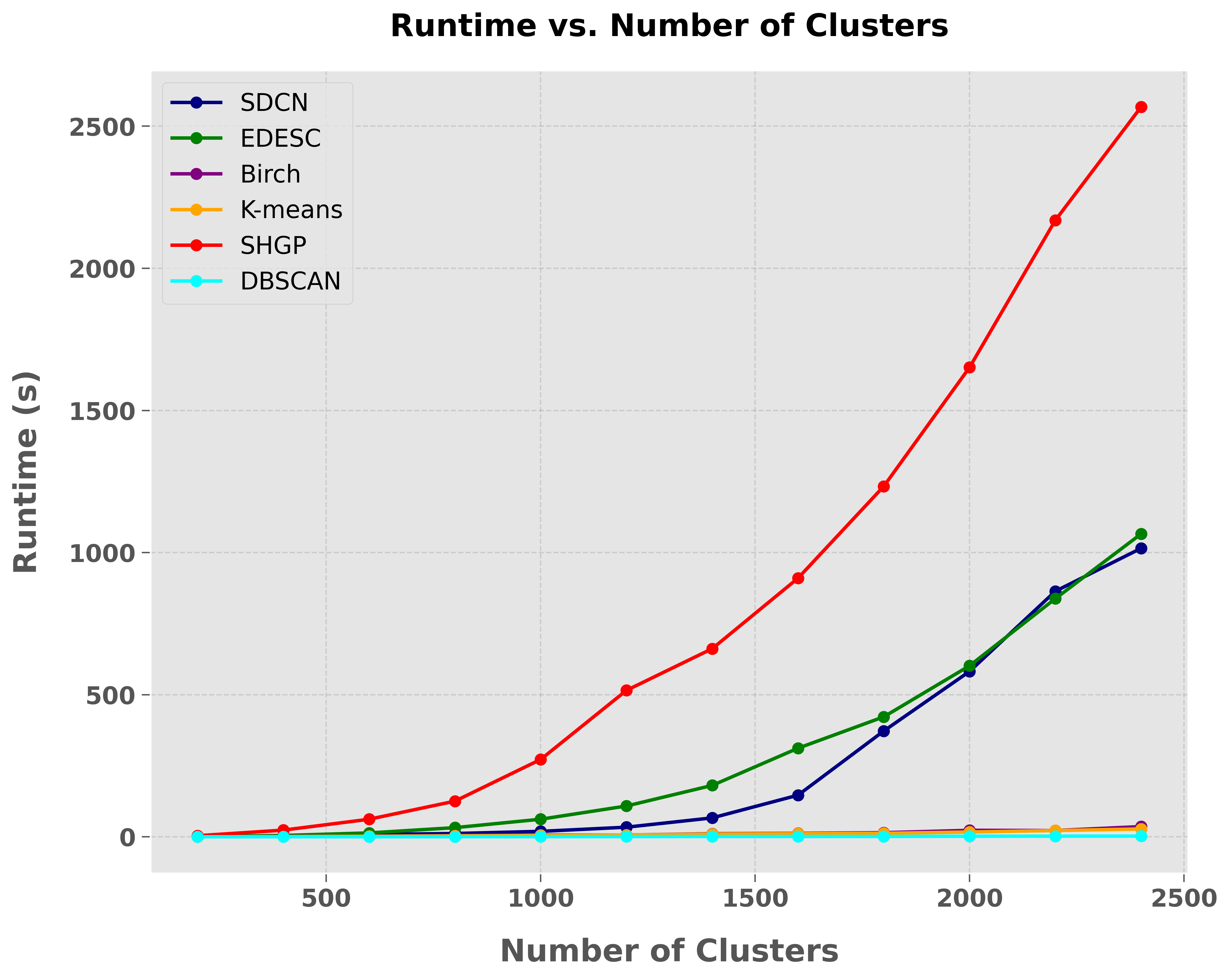}
        \caption{Runtime vs. Number of Clusters $K$.}
        \label{fig:er:runtime-clust}
    \end{subfigure}
    \caption{Runtimes for different numbers of instances and clusters}
    \label{fig:er:runtime}
\end{figure}

The original "Music Brainz 200K" version contains 100,000 GT clusters. As this dataset has a large number of clusters, we use it to investigate the scalability of the DC and SC algorithms, reporting algorithm runtime for varying numbers of clusters and instances (see Figure \ref{fig:er:runtime}). To investigate the impact of the number of instances on performance, we hold $K=200$ as a constant and duplicate the clusters to keep a fixed $K$ for varying numbers of instances.
Figures \ref{fig:er:runtime-inst} shows that the run time of SC methods is significantly lower than that of DC methods, and that run time grows broadly linearly for all methods. Not surprisingly, the DC methods are slower, as DC involves deep neural networks with many hidden layers and parameters, and training of these networks contains the computation of gradients and the updating of network weights, which is computationally expensive. SHGP is several times slower than other DC methods because it uses structural clustering to generate pseudo-labels, and clustering from complex relationships and structures within heterogeneous graphs is time-consuming ~\cite{DBLP:conf/nips/0002GW0XLH22}.

 To investigate the impact of the number of clusters on performance (referred to Figure \ref{fig:er:runtime-clust}), we choose the number of instances corresponding to different values of $K$. We can observe that increasing the number of
clusters significantly impacts the run time of all DC methods.  All DC methods have relatively low run time until about 1000 to 1500 clusters, at which point run times rapidly increase. In SDCN, the computational cost of distance calculations (from each data point $n$ to its cluster centroid) grows linearly when the number of clusters is small, and the runtime is dominated by the number of instances. However, as the number of clusters increases, SDCN needs to do more computation from each data point to each cluster centroid, causing the runtime to grow more than linearly. SHGP uses K-means to get hard cluster labels from low-dimensional embeddings. When we increase $K$, the assignment step (where each data point is assigned to the nearest centroid) takes longer as there are more centroids to compare. Similarly, the process takes longer when the centroid is updated based on its assigned data points because there are more centroids to update. EDESC faces the same issues during the initialization of the subspace with K-means clustering. SC methods have linear runtime growth. K-means updates its centroid (distance calculation to each data point from each centroid) once per iteration and is not impacted by increasing $K$. DBSCAN does not depend on the $K$, but instead, the density of data, leading to the linear runtime. Birch used data points to construct the clustering feature tree, and increasing $K$ does not affect this construction, leading to linear runtime.

Considering the scalability issue of DC methods, we reduce the Music Brainz 20K \cite{DBLP:conf/adbis/SaeediPR17} to  Music Brainz 2K to provide more manageable run times for entity resolution tasks. To ensure the dataset is balanced, we discarded all instances associated with a single cluster, sorting them by cluster-ID in increasing order, and chose the top 2002 instances with 684 clusters. The properties of the datasets evaluated for entity resolution are given in Table \ref{tab:dataset_description}.

 We use schema$+$instance-level data to cluster all records that describe the same real work entity.  Schema-level information is not considered because each record in the Music Brainz 2K dataset contains the same attributes with different descriptions.

Embedding rows with data heterogeneity problems can be challenging, for example, coping with missing attributes for a particular record, the size of the description, and data type ambiguity (for example handling numeric data and multi-word tokens). Consider a scenario of identifying duplicate records with different descriptive patterns \textit{(year:2008,	language:eng)}, \textit{(year:'08',	language:English)}, \textit{(year:,	language:eng)}, and \textit{(year:2008,	length: 24sec)}. The data suffer from several issues, including missing year, year value with numerical and categorical type, same record with different attribute and value abbreviations. Considering these issues, we used EmbDi ~\cite{DBLP:conf/sigmod/CappuzzoPT20} to embed records into the distance matrix, which can be directly input to the DC algorithms. EmbDi ~\cite{DBLP:conf/sigmod/CappuzzoPT20} is based on a tripartite graph with three types of node, specifically value node (representation of unique value), a column node (corresponds to the columns or attribute representation), and a row node (a unique token for each tuple). These nodes are connected in a graph based on the structural information that exists in the dataset. EmbDi adopts random walks between neighboring nodes to capture the local and global structure in the graph, where the length of the random walk and the number of walks per node are user-defined. Column nodes with similar neighborhoods are placed together in the embedding space. EmbDi offers optimizations to handle data heterogeneity problems. We selected only those encodings produced by EmbDi with prefixes (see \cite{DBLP:conf/sigmod/CappuzzoPT20}) $idx\_$, as each token with prefix $idx\_$ represents one tuple.

As SBERT has shown competitive performance on schema inference, we have also applied the pre-trained SBERT model to entity resolution tasks. We computed the SBERT embeddings of each row of the six attributes in the Music Brainz 2K and three attributes in geographic settlements data. 

\subsection{Results and Discussion}
Table \ref{tab:erresult:1} presents the clustering results for entity resolution using Schema$+$Instance-level data. The following can be observed:
(i) \textbf{Running SDCN did not manage to improve the representation compared to AE for any of the datasets.} We observed that SDCN was not further optimizing the representation learned by AE during pre-training as measured by the silhouette score. Due to this, we used the representation of both datasets learned by AE from the pre-training module without considering the clustering loss from SDCN. (ii) {\textbf{Most clustering algorithms produced better results with SBERT than with EmbDi.}  AE with SBERT leads with a 0.26 (for Music Brainz) and 0.24 (for geographic settlements) higher ARI than AE with EmbDi. For Music Brainz data, AE with SBERT obtained 616 more \textit{TP} pairs than AE with EmbDi. For example, one pair, which is \textit{TP} in AE with SBERT and \textit{FN} in AE with EmbDi, is  \textit{(title: 009-Ballade a donner, length: 4m 2sec, artist: Luce Dufault, album: Luce Dufault (1996), year: nan, language: Fre.)} and \textit{(title: Luce Dufault - Ballade Ã donner, length: 242, artist: nan, album: Luce Dufault, year: 96, language: French)}. EmbDi encoded \textit{(length: 242)} as numerical which is given in seconds and \textit{(length: 4m 2sec)} as a string token, whereas SBERT considered both as strings. Similarly, EmbDi did not manage to preserve the contextual information by comparing text with its abbreviations \textit{(language: Fre. vs. language: French)}. 
(iii) \textbf{ The best overall results are with AE for both representations in DC methods.} AE outperforms EDESC on Music Brainz data with 0.08 and 0.34 higher ARI scores with EmbDi and SBERT, respectively, since AE learned features more effectively than those learned during the training with EDESC. ACC shows that AE with SBERT assigns 7\% more samples to the correct clusters in the prediction compared to the number of clusters assigned with EDESC using SBERT. For example, the cosine similarity of two contextually similar SBERT vectors representing \textit{(title: Uriah Heep - Southern Star, length: 266, artist: nan, album: Into the Wild, year: 11, language: English)} and \textit{(title: 0B1-Southern Star, length: 4m 26sec, artist: Heep Uriah, album: Into the Wild (2011), year: nan, language: Eng.)} is 0.78, and should be clustered together with high contextual similarity. However, EDESC placed the two rows in separate clusters compared to AE, which produced the correct clusters. (iv) \textbf{EDESC with SBERT failed to distinguish most of the unary clusters (\textit{TN} in GT) leading it to predict the incorrect number of clusters (668 against 684 GT clusters) compared to AE}. Most EDESC unary clusters have been merged in the prediction, causing a high \textit{FP} rate (EDESC misassigns rows to the same cluster when they should be in different clusters). For example, two instances sharing lexically similar values \textit{(length: 4m 56sec, year: nan, language: Eng.)} and  \textit{(length: 4m 29sec, year: nan, language: Eg.)} belonging to different clusters in GT but obtain high cosine similarity (0.99) in the EDESC latent space with SBERT resulting in a misclassification. (v) \textbf{The original EmbDi representation does not perform especially well, but is improved on by AE and EDESC}. In EmbDi, high similarity scores may be given even where there are few attributes in common. For example, all rows in the largest cluster contain only the common attribute value \textit{(Language: spa.)}, which occurred frequently. The cosine similarity of EmbDi vectors of two records with different values of \textit{(title, length, artist, album, year)} and the same value of \textit{(Language: spa)} is {\it 0.75}, causing the SC algorithms to cluster them together. The representation learned by AE from EmbDi resolved this issue.  (vi) \textbf{SC methods were outperformed by DC methods for both datasets.} Although showing some strength, Birch and K-means do not match the feature learning capabilities of DC methods, reflected in the lower ACC scores than AE (0.19 and 0.10 for Music Brainz with SBERT and 0.05 and 0.32 for geographic settlements with SBERT, respectively).  SHGP, however, failed to produce better clusters than K-means and Birch, with lower ARI scores of 0.14 and 0.16 with EmbDi on geographic settlements data.  Like schema inference, DBSCAN struggles in entity resolution and produces one cluster due to highly similar dense data regions.

\begin{table*}[!htbp]
  \caption{Entity Resolution: clustering results DC (AE, EDESC and SHGP) vs SC (K-means, Birch and DBSCAN) using EmbDi and SBERT on Music Brainz 2K and Geographic Settlements datasets.}
  \label{tab:erresult:1}
\begin{tabular}{cccccccccccccc}
	\toprule  
 &  & \multicolumn{2}{c}{AE} & \multicolumn{2}{c}{EDESC} & \multicolumn{2}{c}{SHGP} & \multicolumn{2}{c}{K-means} & \multicolumn{2}{c}{DBSCAN} & \multicolumn{2}{c}{Birch} \\
 \midrule  
\textbf{Dataset} & \textbf{Metric} & EmbDi & SBERT & EmbDi & SBERT & EmbDi & SBERT & EmbDi & SBERT & EmbDi & SBERT & EmbDi & SBERT \\  
  & $K$ & 684 & 684 & 684 & 668 & 684 & 684 & 684 & 684 & 0 & 1 & 684 & 684 \\  
 Music Brainz & ARI &\textbf{ 0.51} & \textbf{0.77} &  \underline{0.43} & 0.43 & 0.20 & 0.16 & 0.41 & 0.38 & 0.0 & 0.00 & 0.41 &  \underline{0.56} \\  
 & ACC & \textbf{0.71} & \textbf{0.86} &  \underline{0.67} &  \underline{0.79} & 0.51 & 0.48 & 0.65 & 0.67 & 0.002 & 0.004 &  \underline{0.67} & 0.76 \\  
  & $K$ & 786 & 786 & 786 & 786 & 786 & 786 & 786 & 786 & N/A & 1  & 786 & 688\\  
Geographic \\Settlements & ARI & \textbf{0.61} &  \textbf{0.85} &  \underline{0.60} & \underline{0.81}  & 0.43 &  0.72 & 0.57& 0.74& 0.0  & 0.0005  & 0.59 & 0.31 \\  
 & ACC & \underline{0.72} &\textbf{0.91}  & \textbf{0.73} &    \underline{0.89}  & 0.63 & 0.84 &  \underline{0.72}& 0.86  & 0.001& 0.002  & 0.71 & 0.59  \\  
 \bottomrule
\end{tabular}
\end{table*}

\section{Domain Discovery}
\label{sec:dd}

Domain discovery is the process of identifying collections of values that instantiate an application concept.  Discovering domains tends to involve looking for similar collections of values in different dataset columns. Most prior work has used bespoke algorithms~\cite{DBLP:conf/kdd/LiHG17,DBLP:journals/pvldb/OtaMFS20,Pial-22}, but in this section we investigate the use of generic clustering techniques for identifying columns that share domains.

For domain discovery, the clustering problem can be defined as: for a given set of columns $C=\{c_1,c_2,\ c_3\dots c_n\}$ identify every subset $C_s\subseteq C$ that shares a common domain using clustering. To infer a domain from a set of columns, we considered schema-level evidence with pre-trained sentence transformer SBERT and word embedding technique FastText and Schema$+$Instance-level with SBERT and EmbDi \cite{DBLP:conf/sigmod/CappuzzoPT20}.

We used the Di2KG (Camera and Monitor) datasets\footnote{\url{http://di2kg.inf.uniroma3.it/datasets.html}}, which consist of camera and monitor specifications extracted from multiple e-commerce web pages. The datasets are highly heterogeneous in terms of single or multiple sources. For example, synonyms, e.g., \textit{lens}  from \textit{www.cambuy.com.au} and \textit{normalized optical zoom} from \textit{buy.net}, semantically represent the same domain. There are several homonyms, i.e., \textit{screen type} is considered in some sources to represent \textit{screen size}. The properties of the datasets evaluated for domain discovery is presented in Table \ref{tab:dataset_description}. Similar to entity resolution, in some experiments the representation is not well learned in the training of SDCN but by the AE in the pre-training module. Based on the silhouette score, we use the AE instead of SDCN in some domain discovery experiments. The details are given in the hyperparameter setting\footnote{\url{https://github.com/hafizrauf/dc_data-integration}}.

We used three embedding methods for column clustering, considering schema-level and schema$+$instance-level data. To encode column attributes, we used pre-trained models SBERT and FastText as we used in schema inference. To encode columns at schema$+$instance-level, we utilized the Schema Matching (SM) version of EmbDi (Algorithm 5 in EmbDi \cite{DBLP:conf/sigmod/CappuzzoPT20}) and evaluated skip-gram as a learning method with piece-wise smoothing.  In domain discovery, we have a set of columns with cell values that can be represented as a {\it phrase} in SBERT which is trained on diverse text corpora and can capture semantic and syntactic information. Considering this, we used SBERT to encode column headers and values jointly. SBERT processes each column and generates embeddings representing the semantic content of the column headers and values. Subsequently, the embedding for each column is computed by performing a mean operation on the corresponding column header and value embeddings.

\subsection{Results and Discussion}
Table \ref{tab:ddresult:1} shows the clustering results for domain discovery using schema-level data. We observe the following: (i) \textbf{All the clustering algorithms perform quite similarly when considering schema-level data}. This suggests that DC is not significantly improving the representation and indicates that the representations used capture the necessary structure and meaningful differences well enough for SC to group suitable column headers, especially in the Camera dataset.  (ii) \textbf{SHGP outperformed SDCN and EDESC using SBERT with Monitor data}. SHGP obtained an ACC score of 0.03 higher than SDCN and EDESC, in contrast with its performance in domain discovery and entity resolution. SHGP captured more syntactic structures within the column headers and hierarchically divided the graph into various sub-graphs with similar features. For example, attributes \textit{(max resolutions)}, \textit{(resolution)} and \textit{(supported graphics resolutions)} are true positives in GT and SHGP but false negatives in the SDCN prediction. (iii) \textbf{SBERT and FastText with schema-level data are much more similar than in schema inference.} SBERT is leading by 0.03 ARI in SDCN, 0.08 ARI score in EDESC, and 0.05 ARI score in SHGP, a relatively small difference compared to the performance of FastText in SI. This is because, in schema inference, we have long contextual phrases compared to domain discovery. The attribute phrases in the Camera dataset are small, and FastText does not need to consider the order of words to embed, leading to good performance.

\begin{table*}[!htbp]
  \caption{Domain discovery: Schema-level clustering resultsDC (SDCN/AE, EDESC and SHGP) vs SC (K-means, Birch and DBSCAN) using SBERT and FastText on Di2KG (Camera and Monitor) datasets.}
  \label{tab:ddresult:1}
	\begin{tabular}{cccccccccccccc}
	\toprule  
 &  & \multicolumn{2}{c}{SDCN/AE} & \multicolumn{2}{c}{EDESC} & \multicolumn{2}{c}{SHGP} & \multicolumn{2}{c}{K-means} & \multicolumn{2}{c}{DBSCAN} & \multicolumn{2}{c}{Birch} \\  
\textbf{Dataset} & \textbf{Metric} & SBERT  & FastText & SBERT & FastText & SBERT & FastText & SBERT & FastText & SBERT & FastText & SBERT & FastText \\  
\midrule
  & $K$ & 42 & 56 & 56 & 56 & 56 & 56 & 56 & 56 & 49 & 47 & 56 & 44 \\  
Camera & ARI & 0.74 & \textbf{0.71} & \textbf{0.78} &  \underline{0.70} & 0.66 & 0.69 & 0.73 & \textbf{0.71} & 0.73 & 0.35 &  \underline{0.76} & 0.58 \\  
 & ACC & 0.69 & \textbf{0.68} & \textbf{0.74} &  \underline{0.66} & 0.62 & 0.65 & 0.69 &  \underline{0.66} & 0.69 & 0.53 &  \underline{0.70} & 0.62 \\  
  & $K$ & 81 & 81 & 81 & 81 & 81 & 81 & 81 & 81 & 99 & 100 & 81 & 81 \\  
Monitor & ARI &  \underline{0.59} & \textbf{0.57} &  \underline{0.59} &  \underline{0.57} & \textbf{0.59} & 0.54 & 0.57 & 0.55 & 0.27 & 0.30 & 0.52 & 0.54 \\  
 & ACC &  \underline{0.58} & \textbf{0.56 }&  \underline{0.58} & 0.54 & \textbf{0.61} &  \underline{0.55} & 0.57 & 0.54 & 0.50 & 0.51 & 0.54 & \textbf{0.56} \\ 
 \bottomrule 
\end{tabular}
\end{table*}

\begin{table*}[!htbp]
  \caption{Domain discovery: Schema$+$Instance-level clustering results DC (SDCN/AE, EDESC and SHGP) vs SC (K-means, Birch and DBSCAN) using SBERT and EmbDi on Di2KG (Camera and Monitor) datasets.}
  \label{tab:ddresult:2}
	\begin{tabular}{cccccccccccccc} 
	\toprule 
 &  & \multicolumn{2}{c}{SDCN/AE} & \multicolumn{2}{c}{EDESC} & \multicolumn{2}{c}{SHGP} & \multicolumn{2}{c}{K-means} & \multicolumn{2}{c}{DBSCAN} & \multicolumn{2}{c}{Birch} \\  
\textbf{Dataset} & \textbf{Metric} & SBERT  & EmbDi & SBERT & EmbDi & SBERT & EmbDi & SBERT & EmbDi & SBERT & EmbDi & SBERT & EmbDi \\ 
\midrule 
  & $K$ & 51 & 56 & 56 & 56 & 56 & 56 & 56 & 56 & 42 & 1 & 56 & 56 \\  
Camera & ARI & \textbf{0.86} & \textbf{0.13 }&  \underline{0.81} & 0.11 & 0.47 & 0.07 & 0.51 &  \underline{0.12} & -0.005 & 0.02 & 0.78 & 0.03 \\  
  & ACC & \textbf{0.80} & \textbf{0.17} &  \underline{0.78} &  \underline{0.15} & 0.56 & 0.11 & 0.56 &  \underline{0.15} & 0.25 & 0.13 & 0.74 & 0.14 \\  
  & $K$ & 81 & 81 & 81 & 81 & 81 & 81 & 81 & 81 & 87 & 2 & 81 & 81 \\  
Monitor & ARI & \textbf{0.64} & \textbf{0.06} &  \underline{0.62} & \textbf{0.06} & 0.51 &  \underline{0.02} & 0.58 & \textbf{0.06} & 0.05 & 0.002 & 0.60 & 0.04 \\  
 & ACC & \textbf{0.63} & \textbf{0.13} &  \underline{0.62} & \textbf{0.13} & 0.53 &  \underline{0.08} & 0.58 & \textbf{0.13} & 0.38 & 0.06 & 0.61 & \textbf{0.13} \\  
 \bottomrule
\end{tabular}
\end{table*}

\begin{figure}[!htbp]
     \centering
     \begin{subfigure}[b]{0.52\linewidth}
         \centering
         \includegraphics[width=\linewidth]{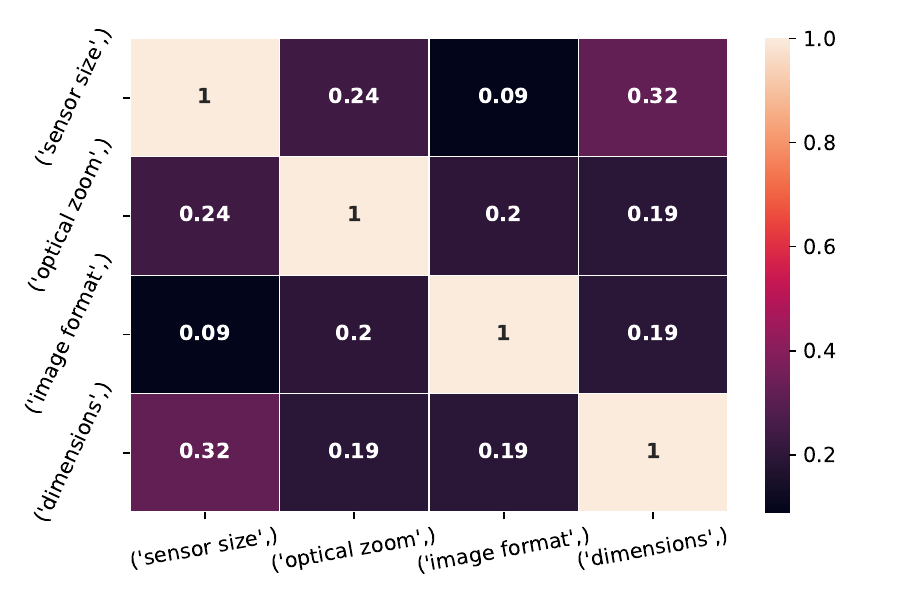}
         \caption{Schema-level}
         \label{fig:dd:1a}
     \end{subfigure}
     \hspace{-0.03\textwidth}
     \begin{subfigure}[b]{0.52\linewidth}
         \centering
         \includegraphics[width=\linewidth]{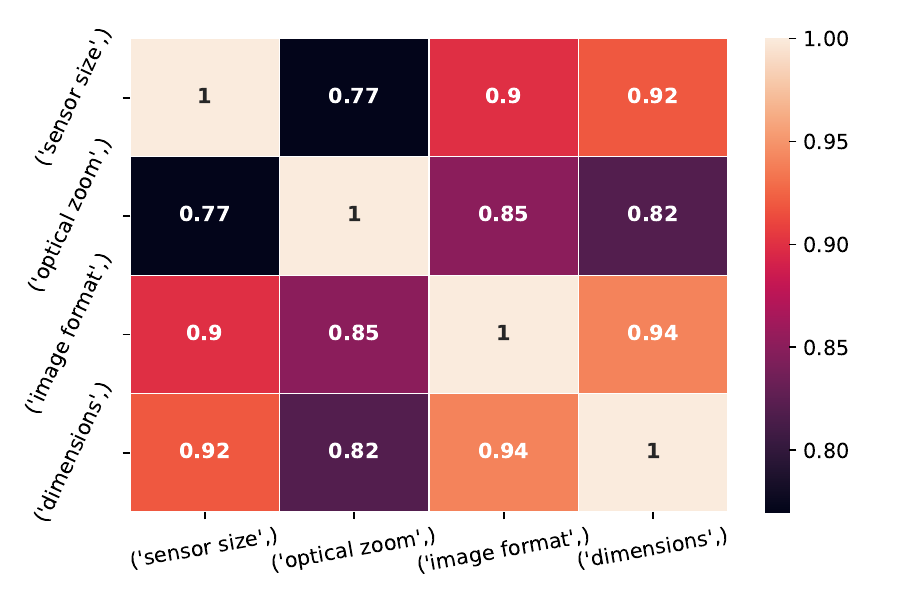}
         \caption{ Schema$+$instance-level}
         \label{fig:dd:1b}
     \end{subfigure}
     \caption{Heat map representation of SBERT (schema-level) (a) and EmbDi (b) with SDCN on Camera data. When instance-level data is added for encoding, we can observe that all true negative cases in (a) are false positives in (b). }
     \label{fig:dd:1}
\end{figure}

Table \ref{tab:ddresult:2} presents the clustering results for domain discovery using schema$+$instance-level data. We observe the following: (i) \textbf{All clustering methods struggled to integrate Schema$+$Instance data with EmbDi and showed much better performance with SBERT on all datasets.} EmbDi failed to produce suitable embeddings for column headers and values because EmbDi emphasizes relationships between columns in a table, which are not especially relevant to domain discovery. In contrast, SBERT considers the textual context for each column header and value and then combines them, ignoring surrounding columns. Furthermore, the performance of EmbDi is also impacted by the syntactic dissimilarity between column headers. For example, two-column attributes in Camera data  \textit{(image size pixels)} and \textit{(max resolution)} are lexically dissimilar with a cosine similarity of 0; however, they can have similar instance values. The largest cluster predicted by EDESC with EmbDi contains 1151 columns that belong to 13 GT domains but represent one domain in the prediction, which shows a high false positive rate. Some examples of domains clustered by EDESC with EmbDi but not in the GT clusters are \textit{(battery type, lens type, battery life, camera type)}. We used heat maps (Figure \ref{fig:dd:1}) to analyze how the distance vectors of columns are similar or dissimilar. We investigate how adding instance-level data affects the representation of columns. For heat map visualization, columns are selected randomly from the predicted clusters of SDCN encoded with SBERT (schema-level) and EmbDi (schema$+$instance-level). Unlike SBERT, EmbDi with SDCN groups those columns that are neither syntactically similar nor belong to the same domain. Adding instance-level data gives rise to a poorer encoding for SDCN with EmbDi. Figure \ref{fig:dd:1a} confirms that different columns that should be in the same cluster are in different clusters. This indicates that the column headers are lexically different and suitable for models pre-trained on large dictionaries. Figure \ref{fig:dd:1b} shows that all the example columns belong to different real-world domains but are still assigned to one cluster. SBERT with SDCN managed to segregate those columns, which are lexically different, from schema-level evidence.  
(ii) \textbf{Combining instance-level data with schema-level data helps in domain discovery but not schema inference for all clustering methods.} In domain discovery, the column headers have high syntactic similarity despite belonging to different domains. Adding more relevant information from column values into the feature space makes the features more different, and clustering methods find criteria to differentiate between clusters. On the other hand, the table attributes in schema inference have high semantic similarity with more similar table values, making the features more similar to each other. Adding table values into feature space may lead to highly overlapping features, which clustering methods find hard to cluster correctly. For example, in domain discovery, the SBERT cosine similarity between two column headers \textit{(headphone outputs)} and \textit{(headphone out)} that belonging to different domains is 0.78, which is relatively high and they are likely to be placed in one cluster; however, when we add instance-level data \textit{(headphone outputs: 1)} and \textit{(headphone out: yes)}, this proides additional information, making the two features less similar.

\section{Conclusions and discussion}
\label{sec:Conclusions and discussion}

This section discusses lessons that can be learned that cut across data integration problems and the study's overall conclusion.
\subsection{Comprehensive Insights}
\label{subsec:Comprehensive Insights}

The following are cross-cutting findings:
\begin{enumerate}

\item SDCN with SBERT consistently outperformed in several problems (particularly for schema inference with schema-level data on web tables, entity resolution on both datasets and domain discovery with schema$+$instance-level on the Camera dataset) compared to other embedding methods. SDCN allows for fine-tuning of SBERT by way of the lower-dimensional latent space of the AE, potentially capturing deeper semantic relationships in sentences. For example, for schema inference, the representation of the two sets of table attributes \textit{(common name, scientific name, family)} and \textit{(species, scientific name, day, high count, total count)} from table \textit{Bird} are well learned by AE when SBERT is fine-tuned compared to the FastText because SBERT considers the context of these attributes, whereas FastText uses sub-word information to gain more syntactic information. The two sets of table attributes are correctly clustered together in SDCN with SBERT but apart with FastText.

\item EDESC performed better clustering when there were a large number of clusters with small cluster cardinality (particularly for schema inference with schema-level on web tables using FastText, domain discovery with schema-level on Camera datasets using SBERT and with schema$+$instance-level on Monitor datasets using EmbDi). When there are a large number of small clusters, those small clusters tend to contain instances with increased discrimination (instances of one cluster are less similar to the instances of others). This occurrence of clusters with prominent distinct features reduces the overlap between different subspaces. In contrast, where there are a small number of large clusters, these tend to have lower inter-cluster distance (fewer other clusters contain clearly distinct instances), increasing the probability of overlapping subspaces in which instances that should be in different clusters are assigned to the same subspace. For example, the GT's mean cluster cardinality for web tables data is 16.5; EDESC with FastText predicted 14 clusters with cardinality below 16.5 compared to SDCN with FastText, which predicted 9, thereby missing more smaller clusters. The same case exists with domain discovery, where the GT’s mean cluster cardinality for the Camera dataset is 340, and EDESC with SBERT produced 43 clusters with cardinality below 340 compared to SDCN with SBERT, which produced 31, again missing more of the smaller clusters.

\item SDCN consistently prioritizes cluster quality over quantity, offering a significant edge over SC methods. SDCN forms fewer clusters (observed for schema inference with schema-level using SBERT on web tables data and domain discovery schema$+$instance-level using SBERT on Camera data), but these clusters are denser and better separated than in SC methods, which even when they produce the same number of clusters as in the GT these are less dense and compact. SDCN learns a compact representation of the data during the clustering process, leading to more cohesive clusters, whereas SC methods use the original feature space, which is not optimized for clustering. A representative example of this phenomenon in domain discovery on schema$+$instance-level Camera data is that SDCN with SBERT formed 42 clusters against 56 GT clusters and yet outperformed Birch and K-means by 0.08 and 0.35 ARI respectively, even though they produced the correct number of clusters.

\item SHGP performs poorly for all problems except when applied to the Monitor dataset, executing domain discovery on schema-level data using SBERT. Since SHGP uses K-means to cluster the embeddings learned by two modules, \textit{Att-LPA} and \textit{Att-HGNN} (referred to SHGP in Section \ref{sec:dc}), this indicates that SBERT embeddings of raw columns are more robust than SHGP embeddings (a fine-tuned version of SBERT using \textit{Att-LPA} and \textit{Att-HGNN}).

\item DBSCAN performed poorly for all experiments in schema inference and entity resolution and predicted a minimal number of clusters, sometimes a singular cluster. We observed that DBSCAN tends to merge distinct clusters into one cluster because all clusters have similar densities. In DBSCAN, a cluster is a dense space region separated by lower-density regions. If all the instances fall in the same density region, it becomes difficult for DBSCAN to differentiate between clusters. We validate this observation using the Kolmogorov-Smirnov (KS) test \cite{an1933sulla, smirnov1948table} to determine the similarity in density distributions between different features. The KS test compares the cumulative distributions of the pairs of instances to determine their differences. The null hypothesis is that the pairs of instances are drawn from the same distribution. We applied the KS test pairwise to all possible pairs of features obtained from SBERT embeddings of web tables data for schema inference considering schema-level data. KS test returns two measures, the K-statistic (smaller value indicates that all features share the same distribution or density) and p-value (smaller value suggests rejecting the null hypothesis). We obtained mean K-statistic = 0.06, indicating that all features represent the same distribution (similar densities), and mean p-value = 0.65, confirming that we cannot reject the null hypothesis.

\end{enumerate}

\subsection{Conclusions}
\label{subsec:Conclusions }

We have investigated the application of DC for {\it schema inference}, {\it entity resolution} and {\it domain discovery}, tasks that cluster tables, rows and columns, respectively. Experiments have explored the use of DC algorithms on these mainstream data management tasks, using a variety of embeddings for complete tables, columns, and rows. Results have been reported comparing three existing DC algorithms with three non-DC algorithms representing different clustering paradigms. The results show that DC algorithms consistently outperform non-DC clustering algorithms for data integration tasks, thus motivating their adoption to cluster tabular datasets or their components. We identified potential future research opportunities by empirical evaluation, which include (i) Exploring distance functions to effectively measure row-to-row, column-to-column and table-to-table similarity in the latent space for deep clustering. (ii) Efficient transformation of dense to sparse matrices before learning the representation. (iii) Exploring different techniques to minimize the effect of large numbers of clusters on deep clustering performance. As the number of clusters grows, the model's complexity increases, and it becomes more likely that some clusters will be very similar, leading to more challenging optimization problems.

\begin{acks}
The authors would like to acknowledge the assistance given by Research IT and the use of the Computational Shared Facility (CSF) at The University of Manchester.
\end{acks}

\bibliographystyle{ACM-Reference-Format}
\bibliography{main}

\end{document}